\documentclass[12pt,preprint]{aastex}
\slugcomment{2010.11 v3}

\shorttitle{4U 0142+61} \shortauthors{H.Tong et al.}

\begin{document}

\title{Non-detection in a Fermi/LAT observation of AXP 4U 0142+61: magnetars?}

\author{H. Tong\altaffilmark{1}
\email{haotong@ihep.ac.cn}}

\author{L. M. Song\altaffilmark{1}
and R. X. Xu\altaffilmark{2}}

\altaffiltext{1}{Institute of High Energy Physics, Chinese Academy
of Sciences} \altaffiltext{2}{School of Physics and State Key
Laboratory of Nuclear Physics and Technology, Peking University}

\begin{abstract}
Significant research in compact stars is currently focused on two
kinds of enigmatic sources: anomalous X-ray pulsars (AXPs) and soft
gamma-ray repeaters (SGRs). Although AXPs and SGRs are popularly thought
to be magnetars, other models (e.g. the accretion model) to understand
the observations can still not be ruled out. It is worth noting that
a non-detection in a Fermi/LAT observation of AXP 4U 0142+61 has been
reported recently by Sasmaz Mus \& Gogus. We propose here that Fermi/LAT
observations may distinguish between the magnetar model and the
accretion model for AXPs and SGRs. We explain how this null observation
of AXP 4U 0142+61 favors the accretion model. Future Fermi/LAT observations
of AXP 1E 1547.0-5408 and AXP 1E 1048.1-5937 are highly recommended.
\end{abstract}


\section{Introduction}

Anomalous X-ray pulsars (AXPs) are pulsar-like objects, whose X-ray
luminosities are in excess of their rotational energy losses while
they show no binary signature, thus acquiring the name ``anomalous''
X-ray pulsars (Mereghetti 2008). AXPs, along with soft gamma-ray
repeaters (SGRs), are candidate  magnetars, neutron stars powered
by strong magnetic field decay (Duncan \& Thompson 1992; Paczynski
1992). Alternative explanations for AXPs and SGRs involve a normal
neutron star accreting from a supernova fallback disk (Alpar 2001; Chatterjee et al. 2000).
It is then a very fundamental question to determine whether AXPs and
SGRs are magnetars or accretion-powered systems. To finally solve
this problem is not only helpful to understand the equation of state
at supra-nuclear densities, but also very meaningful to explain high
energy astrophysical phenomena (Xu 2007).

The magnetar model is prevailing in explaining bursts of AXPs and SGRs (Paczynski 1992; Thomspon \& Duncan 1995).
However, bursting behavior in accretion model is not absolutely impossible (Rothschild et al. 2002; Xu et al. 2006).
It is also possible that the magnetar field ($\sim 10^{14}-10^{15}\,\mathrm{G}$)
responsible for bursts is in higher multipole form
while a normal dipole component ($\sim 10^{12}-10^{13}\,\mathrm{G}$) interacts with the fallback disk (Eksi \& Alpar 2003; Ertan et al. 2007).
Observations in the optical/IR band are informative, e.g. a debris
disk is found around AXP 4U 0142+61 (Wang et al. 2006). The optical/IR observation of 4U 0142+61 can be explained uniformly
in an accretion fallback disk model (Ertan \& Cheng 2004; Ertan et al. 2007).
However, if the disk is passive, a fallback disk is also compitable with the magnetar scenario (Wang et al. 2006).
Therefore observations at other wavelengths are very necessary to understand the real nature of AXPs and SGRs, especially in gamma-rays.

The outer gap model (e.g. Cheng et al. 1986) is very successful, and high energy gamma-ray emissions of AXPs have been calculated and predicted
by Cheng \& Zhang (2001) in the magnetar domain, using the thick outer gap model (Zhang \& Cheng 1997).
The detailed calculations of Cheng \& Zhang (2001) predicted that Fermi/LAT should be able to detect gamma-ray emission of AXPs,
including 4U 0142+61, if they are magnetars. However, a recent Fermi/LAT observation of 4U 0142+61 has been reported,
which shows no detection (Sasmaz Mus \& Gogus 2010). Then there seems a conflict between theory and observation.
While adopting the thick outer gap model
(Zhang \& Cheng 1997), simple calculations show that AXPs are not high energy gamma-ray emitters if they are normal neutron stars accreting from
fallback disks. We suggest that Fermi/LAT observation of AXPs and SGRs can be applied to distinguish
between the magnetar model and the accretion model.
The non-detection of 4U 0142+61 may prefer the accretion model.

In \S 2 we compare theoretical predictions from the magnetar model with Fermi/LAT observation of AXP 4U 0142+61.
Discussions are presented in \S 3.

\section{Theoretical calculations in the magnetar model}

Zhang \& Cheng (1997) developed the thick outer gap model for long period pulsars.
The typical Lorentz factor is determined by equaling energy loss and gain.
The $\gamma-\gamma$ pair production threshold determines the size of the outer gap self-consistently.
If the X-ray photons are provided by surface thermal emission, the size of the outer gap is (eq.(24) in Zhang \& Cheng 1997)
\begin{equation}
 f=4.5 P^{7/6} B_{12}^{-1/2} T_{6}^{-2/3} R_{6}^{-3/2},
\end{equation}
where $P$ is the pulsar rotation period, $B_{12}$ is the stellar magnetic field in units of $10^{12}\,\mathrm{G}$,
$T_{6}$ is the surface temperature in units of $10^6\,\mathrm{K}$, $R_{6}$ is the stellar radius in units of
$10^{6}\,\mathrm{cm}$. Here $f$ should be less than one for outer gap to exist.
In magnetar model for AXPs and SGRs, typical parameters are $P=7\,\mathrm{s}$,
$B=5\times 10^{14}\,\mathrm{G}$, $T=0.5\,\mathrm{keV}$. The stellar radius is chosen as $R=12\,\mathrm{km}$, which is moderate for
realistic equations of state (Lattimer \& Prakash 2007, fig 6 there)
(In Cheng \& Zhang 2001, the stellar radius is chosen as $15\,\mathrm{km}$).
The corresponding outer gap size is then $f=0.46$,
which means that if AXPs and SGRs are magnetars they should be high energy gamma-ray emitters.
On the other hand, if AXPs and SGRs are normal neutron stars
whose (dipolar) magnetic fields are $10^{12}-10^{13}\,\mathrm{G}$ (Alpar 2001; Chatterjee et al. 2000),
the corresponding outer gap size is $f=3-10$. Therefore if AXPs and SGRs are normal neutron stars accreting from fallback disks,
they will not radiate high energy gamma-rays\footnote{Ertan \& Cheng (2004) argued that accretion-powered system can also emit high energy gamma-rays
if the inner disk rotates faster than the neutron star. However, this criterion cannot be matched
for the debris disk around 4U 0142+61 either as a passive disk (Wang et al. 2006) or as a gaseous accretion disk
(Ertan et al. 2007).}. Thus Fermi/LAT observations of AXPs and SGRs can be helpful to distinguish between the magnetar model and the accretion model.

Sasmaz Mus \& Gogus (2010) reported Fermi/LAT observation of AXP 4U 0142+61.
With an exposure time of $31.7\,\mathrm{Ms}$, they find no detection of high energy gamma-ray
emission from 4U 0142+61 in both $0.2-1\,\mathrm{GeV}$ and $1-10\,\mathrm{GeV}$ band.
Observational upper limits and theoretical calculations in the magnetar model are shown in figure 1.
For 4U 0142+61, its parameters are $P=8.688\,\mathrm{s}$, $B=2.6\times 10^{14}\,\mathrm{G}$, $T=0.395\,\mathrm{keV}$
(from the McGill AXP/SGR catalog\footnote{http://www.physics.mcgill.ca/$\sim$pulsar/magnetar/main.html}).
The magnetic field is calculated from $B=6.4\times10^{19}\sqrt{P\dot{P}}$, which is 2 times larger than usually reported since
the polar magnetic field is more important in the case of pulsar radiation (Shapiro \& Teukolsky 1983).
The size of the outer gap for 4U 0142+61 is $f=0.96$.
The distance $d=2.5\,\mathrm{kpc}$ and solid angle $\Delta \Omega=1$ are used during the calculation. From figure 1,
the observational upper limits are below the theoretical calculations for large inclination angles ($60^{\circ}$, $75^{\circ}$)
or when the inner boundary of outer gap can extend to 10 stellar radii (Hirotani et al. 2003; Hirotani \& Shibata 2001).

\begin{figure}[!t]
 \centering
 \includegraphics[width=0.75\textwidth]{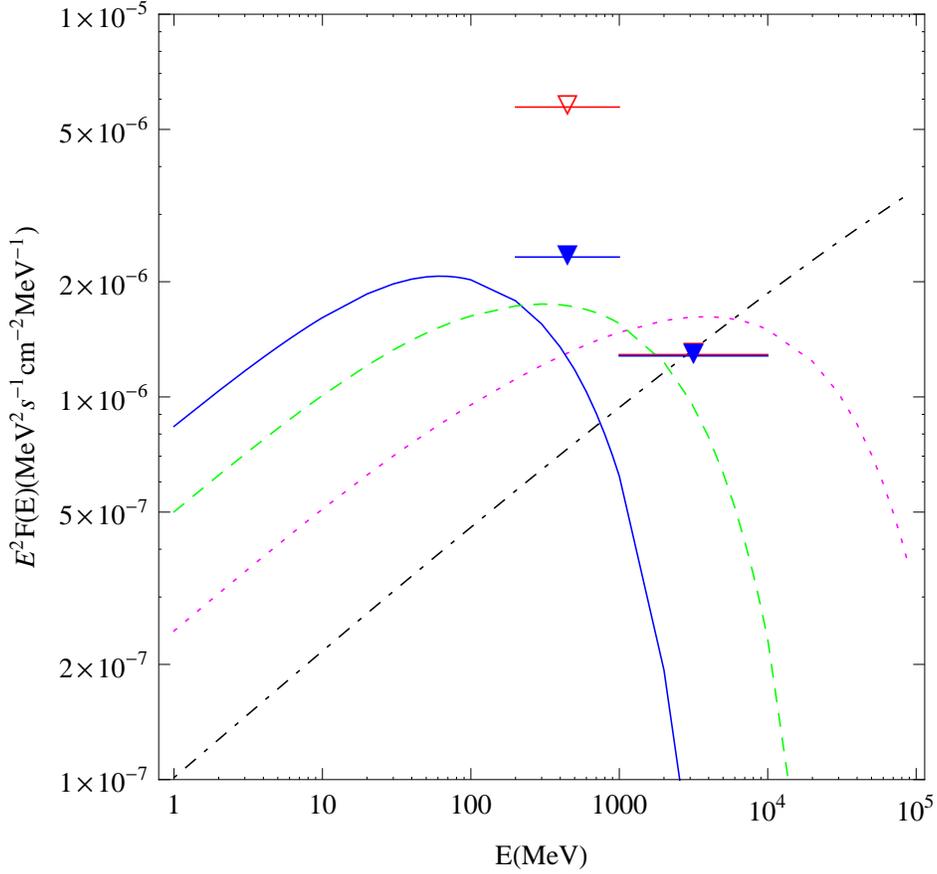}
 \caption{Fermi/LAT upper limits of AXP 4U 0142+61 compared with outer gap calculations in the magnetar domain.
The solid, dashed, and dotted lines are for inclination angle $45^{\circ}$, $60^{\circ}$, $75^{\circ}$ respectively
(Zhang \& Cheng 1997; Cheng \& Zhang 2001).
The dotdashed line takes into consideration that the inner boundary of outer gap may extend to 10 stellar radii
(Hirotani et al. 2003; Hirotani \& Shibata 2001).
The empty down triangle and filled down triangle are Fermi/LAT upper limits ($0.2-1\,\mathrm{GeV}$ and $1-10\,\mathrm{GeV}$)
from $2^{\circ}$ and $15^{\circ}$ extraction region respectively
(Sasmaz Mus \& Gogus 2010). The upper limits in $1-10\,\mathrm{GeV}$ are nearly coincide.}
\end{figure}

Possible reasons why Fermi/LAT has not seen the expected high energy gamma-rays from AXP 4U 0142+61 are:
\begin{enumerate}
 \item Its radius is smaller than $12\,\mathrm{km}$;
 \item Its distance is much larger than $2.5\,\mathrm{kpc}$;
 \item The inclination angle is small, e.g. $45^{\circ}$;
 \item Beaming of gamma-ray radiation;
 \item The radiated high energy gamma-ray photons are absorbed due to internal or external matter.
\end{enumerate}
For order of magnitude estimations, the neutron star radius is often taken as $10\,\mathrm{km}$.
However, for realistic equations of state, this choice corresponds to a soft equation of state (Lattimer \& Prakash 2007).
For a stiff equation of state
the radius can be as large as $15\,\mathrm{km}$. A radius of $12\,\mathrm{km}$ is a moderate choice (Lattimer \& Prakash 2007, fig 6 there).
Neutron star equation of state studies (e.g. Tsuruta 2006) also prefer medium to stiff equations of state.

Figure 2 shows the model calculations for AXP 4U 0142+61 when the distance is 2 times larger, i.e. $5\,\mathrm{kpc}$, along with
Fermi/LAT sensitivity curve for $5\sigma$ detection (Atwood et al. 2009). Even when the distance is 2 times larger than we presently
employed, Fermi/LAT should also be able to detect the expected gamma-ray emission of 4U 0142+61. Also in figure 2, when the inclination angle is small,
e.g. $45^{\circ}$, its high energy radiation is decreased along with an increase in the low energy part
(cf. fig 4 in Cheng \& Zhang 2001).
Therefore if the inclination angle is small, although Fermi/LAT could not detect 4U 0142+61 in $(1-10)\,\mathrm{GeV}$ band,
it could detect 4U 0142+61 in $(0.1-1)\,\mathrm{GeV}$ and lower energy band. In Cheng \& Zhang (2001), the inclination angle determines the
inner boundary of the outer gap. Recent modeling indicates that the inner boundary may extend to 10 stellar radii
(Hirotani et al. 2003; Hirotani \& Shibata 2001). Employing this assumption, the corresponding model calculations are shown in
figure 1 and 2.

According to Cheng \& Zhang (2001) and references therein,
the solid angle for known gamma-ray pulsars ranges from $0.5-2.5$. Recent Fermi observations of gamma-ray pulsars also show a relatively
broad pulse profile (Ray \& Parkinson 2010). Therefore the beaming of gamma-ray radiation is not the key factor obscuring our observation
of gamma-ray emissions and this problem can be cleared with future Fermi/LAT observations of more AXPs and SGRs.

The magnetic field at the inner boundary of outer gap is $2.6\times 10^5\,\mathrm{G}$ for inclination angle $75^{\circ}$
(or $2.6\times 10^{11}\,\mathrm{G}$ when the inner boundary is chosen as 10 stellar radii). The absorption of high energy photons
is not significant at the inner boundary due the weakness of the magnetic field (Ruderman \& Sutherland 1975). For AXP 4U 0142+61,
it has a debris disk whose photon energy is typically $0.1-1\,\mathrm{eV}$ (Wang et al. 2006). The $\gamma-\gamma$ absorption
is negligible for GeV photons (Zhang \& Cheng 1997).

In conclusion, based on the thick outer gap model (Zhang \& Cheng 1997), for a variety of the parameter space in magnetar model,
Fermi/LAT should be able to detect the expected high energy gamma-ray emission from AXP 4U 0142+61.
This is in conflict with Sasmaz Mus \& Gogus (2010).

\begin{figure}[!t]
 \centering
 \includegraphics[width=0.75\textwidth]{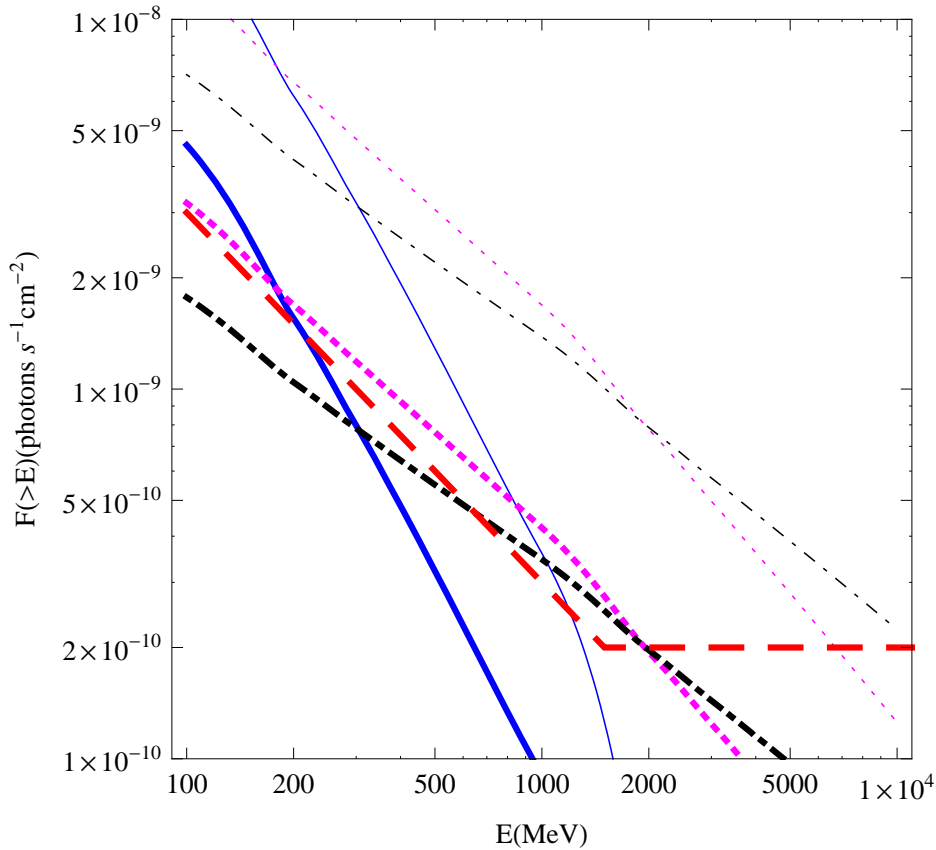}
 \caption{Fermi/LAT integral sensitivity curve and model calculations for AXP 4U 0142+61.
The solid, dotted, and dotdashed lines are the same as those in figure 1, except that
the integral flux is shown instead of differential flux. The corresponding thick lines  are model calculations
when the distance is 2 times larger, i.e. $5\,\mathrm{kpc}$. The thick dashed line is the Fermi/LAT sensitivity curve (Atwood et al. 2009).}
\end{figure}

\section{Discussions}

At the beginning of \S 2, we show that AXPs are not high energy gamma-ray emitters ($f$ larger than 1)
if they are normal neutron stars accreting from fallback disks. Therefore the non-detection
in a Fermi/LAT observation of AXP 4U 0142+61 can be naturally explained in the accretion model
for AXPs. The spectra energy distribution of 4U 0142+61 indicates an energy break at about $1\,\mathrm{MeV}$ (Sasmaz Mus \& Gogus 2010).
If hard X-ray emission of 4U 0142+61 originates from near the stellar surface,
the energy break is also at $1\,\mathrm{MeV}$ for a normal neutron star (Zhang \& Cheng 1997).
Of course the detailed origin of AXP hard X-ray emission needs further studies.

In the accretion model for AXPs (Alpar 2001; Chatterjee et al. 2000) (also for SGRs, if they are indeed one population),
the long period of AXPs is due to disk braking in the propeller phase. They are now
X-ray luminous since they have entered the accretion phase. The bursts of AXPs and SGRs may due to
accretion induced quakes (AIQs) (Xu et al. 2006; Xu 2007), or quakes and plate tectonics of neutron stars
(Rothschild et al. 2002). The accretion induced quake model of Xu et al. (2006) provides a link between
persistent emission and bursts. A hybrid model is also possible which the magnetar field is in
higher multipole form and the spin down is governed by a normal dipole component interacting with a fallback disk (Eksi \& Alpar 2003).
The recently reported low magnetic field SGR (SGR 0418+5927 with $B_{\mathrm{dipole}}<7.5\times 10^{12}\,\mathrm{G}$, Rea et al. 2010)
is consistent with the accretion model.

For AXP 4U 0142+61, as noted in section 2, it will not
emit high energy gamma-rays even if it is a magnetar, when its radius is $10\,\mathrm{km}$ instead of $12\,\mathrm{km}$.
Therefore future Fermi/LAT observations of more AXPs and SGRs are very necessary.
Outer gap predictions in the magnetar domain for other AXPs and SGRs are shown in figure 3.
Model calculations for three AXPs and one SGR are shown, using observational parameters from the McGill AXP/SGR online catalog.
For gamma-ray luminous and nearby sources, model calculations of AXP 1E 1547.0-5408 and AXP 1E 1048.1-5937 are well above the
Fermi/LAT sensitivity curve. Therefore future Fermi/LAT observations of these two sources are highly recommended.
Among other AXPs, some are not supposed to be high-energy gamma-ray emitters ($f$ larger than 1),
some have relatively low gamma-ray luminosities as shown for AXP XTE J1810-197 in figure 3,
some lies too far away from us. For the two candidate high energy gamma-ray emitting SGRs, SGR 1806-20 and SGR 1900+14,
they are too far away to be detected by Fermi/LAT, as shown for SGR 1806-20 in figure 3.

In conclusion, based on the thick outer gap model (Zhang \& Cheng 1997), the non-detection in a Fermi/LAT observation of
AXP 4U 0142+61 may prefer the accretion model. Future Fermi/LAT observations of AXP 1E 1547.0-5408 and AXP 1E 1048.1-5937
will help us make clear whether they are magnetars or not\footnote{During the submission of this paper, the Fermi-LAT collaboration
have published their observations for all known AXPs and SGRs (Abdo et al. 2010), where still no significant detection is reported.
This result is in favor of our conclusions.}.

\begin{figure}[!t]
 \centering
 \includegraphics[width=0.75\textwidth]{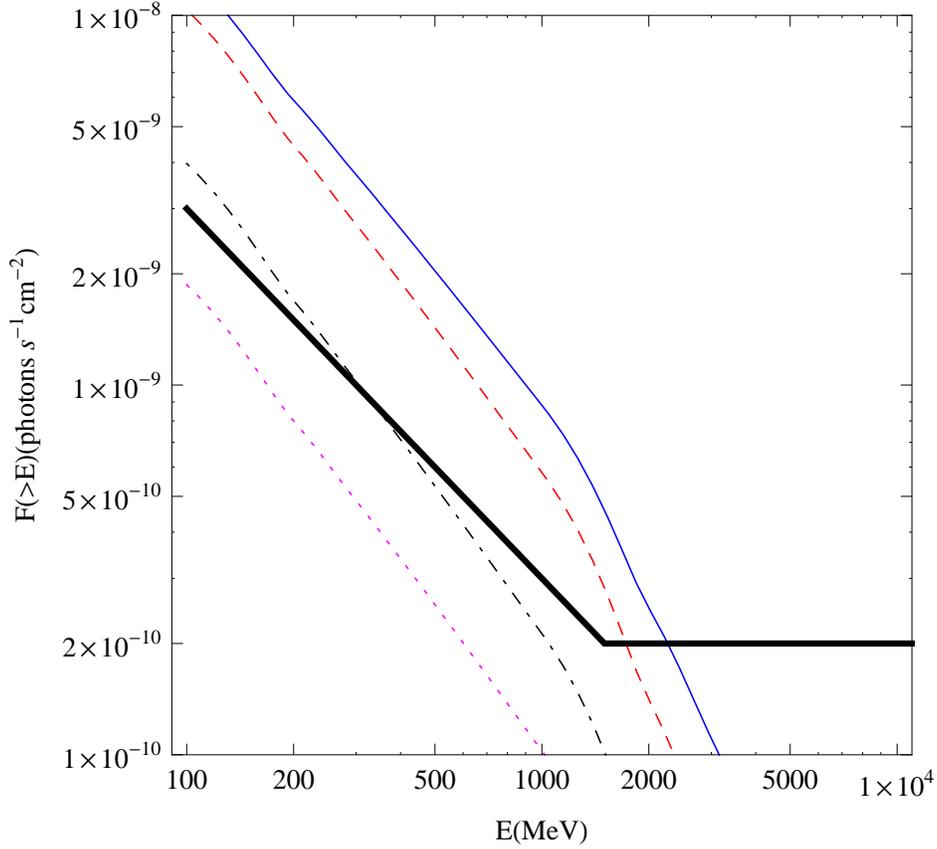}
 \caption{Model calculations for other AXPs and SGRs (Zhang \& Cheng 1997; Cheng \& Zhang 2001).
The inclination angle is chosen as $60^{\circ}$ and star raidus $10\,\mathrm{km}$.
The solid, dashed, dotdashed and dotted lines are for AXP 1E 1547.0-5408, AXP 1E 1048.1-5937, AXP XTE J1810-197 and SGR 1806-20, respectively.
The thick line is the Fermi/LAT sensitivity curve (Atwood et al. 2009).}
\end{figure}

\section*{Acknowledgments}
The authors would like to thank Dr. Stephen Justham for help in the English of this paper.
This work is supported by the National Natural Science Foundation
of China (Grant Nos. 10935001, 10973002), and the National Basic Research Program of China
(Grant No. 2009CB824800)

\end{document}